\documentclass[twocolumn]{aastex61}

\def\JB{{\rm Jy~beam^{-1}}}
\def\mJB{{\rm mJy~beam^{-1}}}

\def\mJ{{\rm mJy}}
\def\kms{{\rm km~s^{-1}}}
\def\Ms{M_{\sun}}
\def\Ls{L_{\sun}}
\def\Rs{R_{\sun}}

\received{}
\revised{\today}
\accepted{}
\submitjournal{ApJL}

\shorttitle{the protostar SMM11 in Serpens Main}
\shortauthors{Aso et al.}

\begin{document}

\title{ALMA observations of SMM11 reveal an extremely young protostar in Serpens Main cluster}

\correspondingauthor{Yusuke Aso}
\email{yaso@asiaa.sinica.edu.tw}

\author[0000-0002-8238-7709]{Yusuke Aso}
\affil{Academia Sinica Institute of Astronomy and Astrophysics P.O. Box 23-141, Taipei 106, Taiwan}
\affil{Subaru Telescope, National Astronomical Observatory of Japan 650 North A'ohoku Place, Hilo, HI 96720, USA}

\author{Nagayoshi Ohashi}
\affil{Academia Sinica Institute of Astronomy and Astrophysics P.O. Box 23-141, Taipei 106, Taiwan}
\affil{Subaru Telescope, National Astronomical Observatory of Japan 650 North A'ohoku Place, Hilo, HI 96720, USA}

\author{Yuri Aikawa}
\affil{Department of Astronomy, Graduate School of Science, The University of Tokyo, 7-3-1 Hongo, Bunkyo-ku, Tokyo 113-0033, Japan}

\author{Masahiro N. Machida}
\affil{Department of Earth and Planetary Sciences, Faculty of Sciences Kyushu University, Fukuoka 812-8581, Japan}

\author{Kazuya Saigo}
\affil{Chile Observatory, National Astronomical Observatory of Japan, Osawa 2-21-1, Mitaka, Tokyo 181-8588, Japan}

\author{Masao Saito}
\affil{Nobeyama Radio Observatory, Nobeyama, Minamimaki, Minamisaku, Nagano 384-1305, Japan}
\affil{SOKENDAI, Department of Astronomical Science, Graduate University for Advanced Studies}

\author{Shigehisa Takakuwa}
\affil{Department of Physics and Astronomy, Graduate School of Science and Engineering, Kagoshima University, 1-21-35 Korimoto, Kagoshima, Kagoshima 890-0065, Japan}
\affil{Academia Sinica Institute of Astronomy and Astrophysics P.O. Box 23-141, Taipei 106, Taiwan}

\author{Kengo Tomida}
\affil{Department of Earth and Space Science, Osaka University, Toyonaka, Osaka 560-0043, Japan}

\author{Kohji Tomisaka}
\affil{National Astronomical Observatory of Japan, Osawa, 2-21-1, Mitaka, Tokyo 181-8588, Japan}

\author{Hsi-Wei Yen}
\affil{European Southern Observatory, Karl-Schwarzschild-Str. 2, D-85748 Garching, Germany}

\author{Jonathan P. Williams}
\affil{Institute for Astronomy, University of Hawaii at Manoa, Honolulu, Hawaii, USA}



\begin{abstract}

We report the discovery of an extremely young protostar, SMM11, located in the associated submillimeter condensation in the Serpens Main cluster using the Atacama Large Millimeter/submillimeter Array (ALMA) during its Cycle 3 at 1.3 mm
and an angular resolution of $\sim 0\farcs 5\sim 210$ AU. SMM11 is a Class 0 protostar without any counterpart at 70 $\micron$ or shorter wavelengths.
The ALMA observations show 1.3 mm continuum emission associated with a collimated $^{12}$CO bipolar outflow. $Spitzer$ and $Herschel$ data show that SMM11 is extremely cold ($T_{\rm bol}=$26 K) and faint ($L_{\rm bol}\lesssim 0.9\Ls$). We estimate the inclination angle of the outflow to be $\sim 80 \arcdeg $, almost parallel to the plane of the sky, from simple fitting using wind-driven-shell model. The continuum visibilities consist of Gaussian and power-law components, suggesting a spherical envelope with a radius of $\sim 600$ AU around the protostar. The estimated low C$^{18}$O abundance, $X$(C$^{18}$O)=1.5-3$\times 10^{-10}$, is also consistent with its youth. The high outflow velocity, a few 10 $\kms$ at a few 1000 AU, is much higher than theoretical simulations of first hydrostatic cores and we suggest that SMM11 is a transitional object right after the second collapse of the first core.
\end{abstract}




\section{INTRODUCTION} \label{ch4:sec:intro}
Since the first hydrostatic core (FHSC) phase was theoretically predicted by \citet{la1969}, several FHSC ``candidates'' have been observationally identified based either on the lack of $Spizter$ infrared detections indicating low temperature (10-30 K), or slow outflows ($\lesssim 5\ \kms$), and/or chemical evolution: Cha-MMS1 \citep{be2006}, L1448 IRS2E \citep{ch2010}, L1451-mm \citep{pi2011}, Per-Bolo58 \citep{du2011} CB 17 MMS \citep{ch2012}, B1-bS \citep{hi.li2014}, and B1-bN \citep{hi.li2014}. Nevertheless, none of these sources can be unambiguously characterized as true FHSCs partly because the theoretically-predicted parameter space of first cores is wide-spread; mass, lifetime, internal luminosity\footnote{Internal luminosity does not include the luminosity due to external heating by interstellar radiation field and envelopes.} and radius of $0.01-0.1\ \Ms$, 500-5$\times 10^{4}$ yr, $10^{-4}-0.1\ \Ls$, and 5-100 AU, respectively \citep{bo.yo1995,ma1998,om2007,co2012,sa.to2006,sa2008,to2010}. In addition, observations have not strongly constrained properties such as the rotation and mass accretion rate in the early phases of star formation. Simulations of rotating FHSCs suggest that those FHSCs \citep{ba2011,ma.ma2011} can transform into Keplerian disks around protostars \citep{aso2015}. Observationally identifying FHSCs is, therefore, important to understand disk formation as well as star formation.  

SMM11 is one of the submillimeter continuum condensations in the Serpens Main cluster forming region \citep[$d=429$ pc;][]{dz2011}, identified in JCMT observations \citep{da1999} and located at the southern edge of Serpens Main. A star-forming core is identified in CARMA observations \citep{leKI2014} in 3 mm at an angular resolution of $\sim 8\arcsec$, associated with a bipolar HCN outflow with a length of $\sim 1000$ AU and a velocity of $\sim 6\ \kms$. They revealed that the core is located at an intersection of two filaments. Its core radius and mass were estimated from the CARMA observations to be $\sim 900$ AU and 1.35 $\Ms$, respectively. However it was not detected by $Spitzer$ \citep{en2009,ev2009,du2015}, in X ray \citep{gi2007}, nor in 6 cm \citep{or2015}. The latter two trace magnetic activities due to convection in a second core, i.e., protostar.

In this paper we report ALMA Cycle 3 observations toward a protostar, SMM11, in the associated submillimeter continuum condensation in Serpens Main in $^{12}$CO $J=2-1$ line, C$^{18}$O $J=2-1$ line, and 1.3 mm continuum, which reveal that SMM11 is in an extremely early phase right after the second collapse.

\section{ALMA OBSERVATIONS} \label{ch4:sec:obs}
We observed five fields in Serpens Main cluster, their locations based on unpublished SMA data of a mosaicking survey carried out in 2010, using ALMA at its Cycle 3 stage on 2016 May 19 and 21. Observations toward the dusty core of SMM11 is reported in this paper and those of the other four regions will be reported in future papers. 
On-source observing time for SMM11 is 4.5 and 9.0 min in the first and the second days, respectively. The numbers of antenna were 37 and 39 in the first and the second days, respectively, and the antenna configuration of the second day was more extended than that of the first day. Any emission beyond $8\farcs0\sim 3400$ AU was resolved out by $\gtrsim 50\%$ with the antenna configuration \citep{wi.we1994}.
Spectral windows for $^{12}$CO $(J=2-1)$ and C$^{18}$O ($J=2-1$) line emissions have 3840 and 1920 channels covering 117 and 59 MHz band width, respectively, at a frequency resolution of $\sim 30.5$ kHz. In this paper, 16 and 2 channels are binned for $^{12}$CO and C$^{18}$O lines and the resulting velocity resolutions are 0.63 and $0.083\ \kms$, respectively. Two other spectral windows cover 216-218 GHz and 232-234 GHz, and used to measure the continuum emission.

All the imaging processes were performed with the Common Astronomical Software Applications (CASA). The visibilities were Fourier transformed and CLEANed with Briggs weighting, a robust parameter of 0.0, and a threshold of 3$\sigma$. Multi-scale CLEAN was used for the line maps to converge CLEAN, where CLEAN components were point sources or $\sim 1\farcs5$ Gaussian sources.

We also performed self-calibration for the continuum data using tasks in CASA ($clean$, $gaincal$, and $applycal$). This improved the rms noise level of the continuum maps by a factor of $\sim 2$. These  calibration tables for the continuum observations were then applied to the line data. The noise level of the line maps were measured in emission-free channels. The parameters of our observations are summarized in Table \ref{ch4:tab:obs}.

\begin{deluxetable*}{cccc}
\tablecaption{Summary of the ALMA observational parameters \label{ch4:tab:obs}}
\tablehead{
\colhead{Date} & \multicolumn{3}{c}{2016.May.19, 21}\\
\colhead{Projected baseline length} & \multicolumn{3}{c}{15 - 613 m $\sim$ 11 - 460 k$\lambda$}\\
\colhead{Primary beam} & \multicolumn{3}{c}{$27\arcsec$}\\
\colhead{Passband calibrator} & \multicolumn{3}{c}{J1751$+$0939}\\
\colhead{Flux calibrator} & \multicolumn{3}{c}{Titan}\\
\colhead{Gain calibrator} & \multicolumn{3}{c}{J1830$+$0619 (470 mJy), J1824$+$0119 (79 mJy)}\\
\colhead{Coordinate center (J2000)} & \multicolumn{3}{c}{$18^{\rm h}30^{\rm m}$00\fs 38, $1^{\circ}11\arcmin 44\farcs 55$}\\
}
\startdata
 & Continuum & $^{12}$CO ($J=2-1$) & C$^{18}$O ($J=2-1)$\\
\hline
Frequency (GHz) & 225 & 230.538000 & 219.560358\\
Bandwidth/velocity resolution & 4 GHz & $0.63\ \kms$ & $0.083\ \kms$\\
Beam (P.A.) & $0\farcs 58\times 0\farcs47\ (-86\arcdeg)$ & $0\farcs 62\times 0\farcs 51\ (-83\arcdeg)$ & $0\farcs 65\times 0\farcs 52\ (-85\arcdeg)$\\
rms noise level ($\mJB$) & 0.074 & 5.0 & 11\\
\enddata
\end{deluxetable*}

\section{RESULTS}
\label{ch4:sec:results}
\begin{figure*}[ht!]
\epsscale{1.2}
\plotone{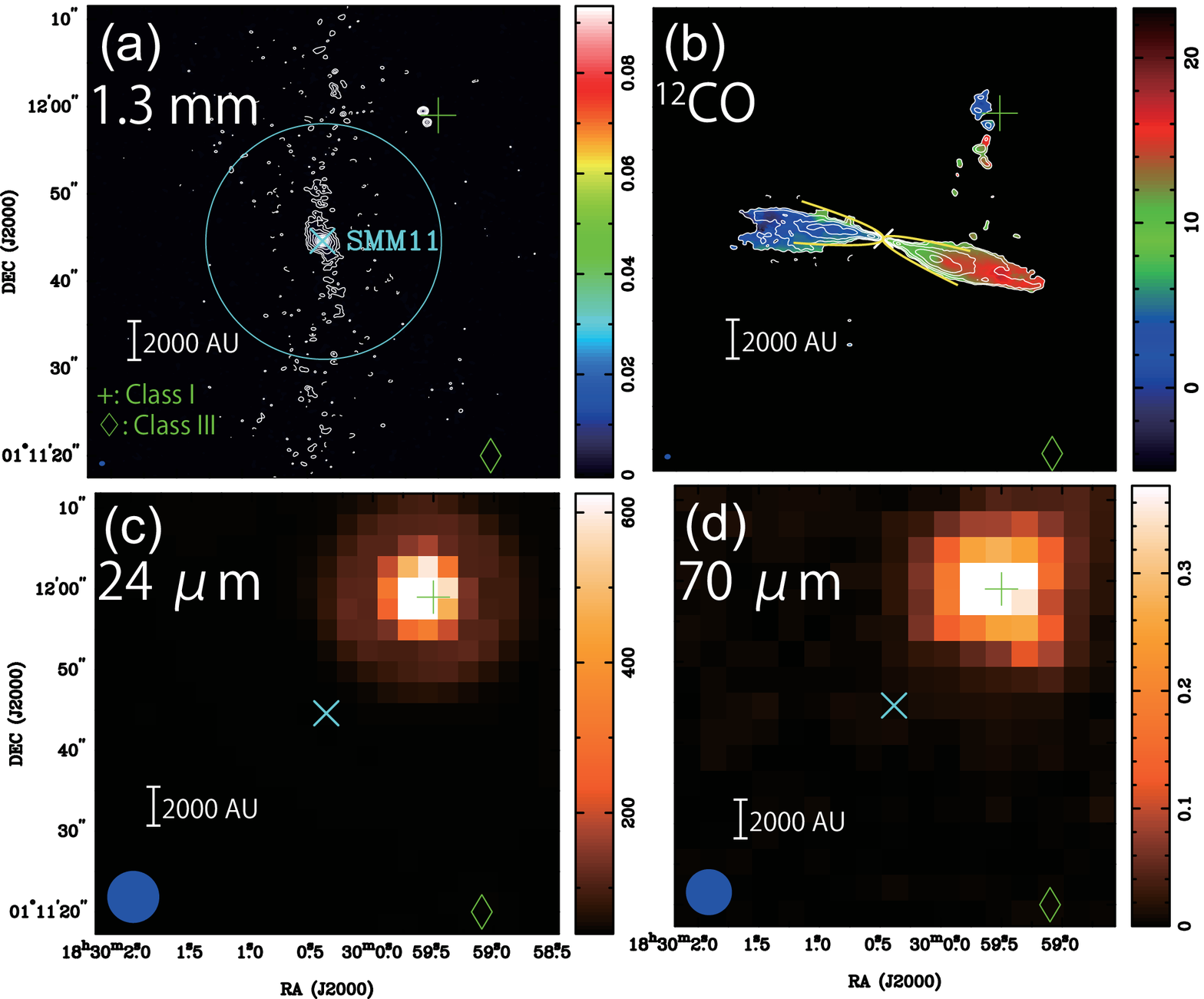}
\caption{SMM11 images in line and continuum emissions. (a) 1.3 mm continuum. Contour levels are $3,6,12,24...\times \sigma$. Green marks indicate YSO positions \citep{du2015}. Cyan circle indicates the ALMA primary beam. (b) Integrated intensity (contours) and mean velocity (color) maps of the $^{12}$CO emission, integrated from $V_{\rm LSR}=-5$ to $23\ \kms$ except for 8-$10\ \kms$ where $^{12}$CO emission is affected by self-absorption and missing flux. Contour levels are $5,10,20,40...\times \sigma$, where $1\sigma$ corresponds to $24\ \mJB~\kms$. Yellow curves show the best-fit parabolic models (see Section \ref{ch4:sec:flow} in more detail). (c) $Spitzer$ 24 $\micron$ image in ${\rm Jy}~{\rm sr}^{-1}$. (d) $Herschel$ 70 $\micron$ image in ${\rm Jy}~{\rm pixel}^{-1}$, where 1 pixel is $3\farcs2 \times 3\farcs2$. Blue filled ellipses at each bottom-left corner denote ALMA synthesized beams or PSFs. X marks denote the peak position in 1.3 mm.
\label{ch4:fig:11c}}
\end{figure*}

Figure \ref{ch4:fig:11c} shows 1.3 mm and $^{12}$CO images of the SMM11 region as well as 24 $\micron$ and 70 $\micron$ images for comparison. Strong compact emission was confirmed in 1.3 mm at the mapping center. We call this source SMM11 in this paper. The continuum emission shows a compact component and an S-shaped component surrounding the compact component. The emission is also extended in the north-south direction on a $\sim 10\arcsec$ scale, which is a part of the filaments. Two-dimensional Gaussian fitting to the 1.3 mm image provides an approximately circular deconvolved size, 160 AU $\times$ 130 AU (P.A.=80$^{\circ}$) at the distance of Serpens Main. The derived total flux density, $164\pm 1\ \mJ$, corresponds to a total mass of 0.09-0.27 $\Ms$ assuming a dust temperature, $T_{\rm dust}=$20-50 K \citep{ha2015}, dust opacity, $(\kappa_{850 \mu {\rm m}},\beta)=(0.035\ {\rm cm}^{2}\,{\rm g}^{-1},1)$ \citep{an.wi2005}, and a gas-to-dust mass ratio of 100. In the northwest of SMM11, two other compact emissions separated by $\gtrsim 1\arcsec \sim 430$ AU were detected beyond the primary beam, indicating the presence of a binary inside the apparently single Class I source (SSTc2d J182959.5$+$011159) seen in the infrared images (Figure \ref{ch4:fig:11c}c-d). The total flux density of the binary at 1.3 mm is $\sim 8\ \mJ$ before primary beam correction. The $^{12}$CO emission traces a bipolar outflow with a size of $\sim6400 \times 1300$ AU. The $^{12}$CO emission also traces another bipolar outflow associated with the binary in the northwest of SMM11.

\begin{figure}[ht!]
\epsscale{0.9}
\plotone{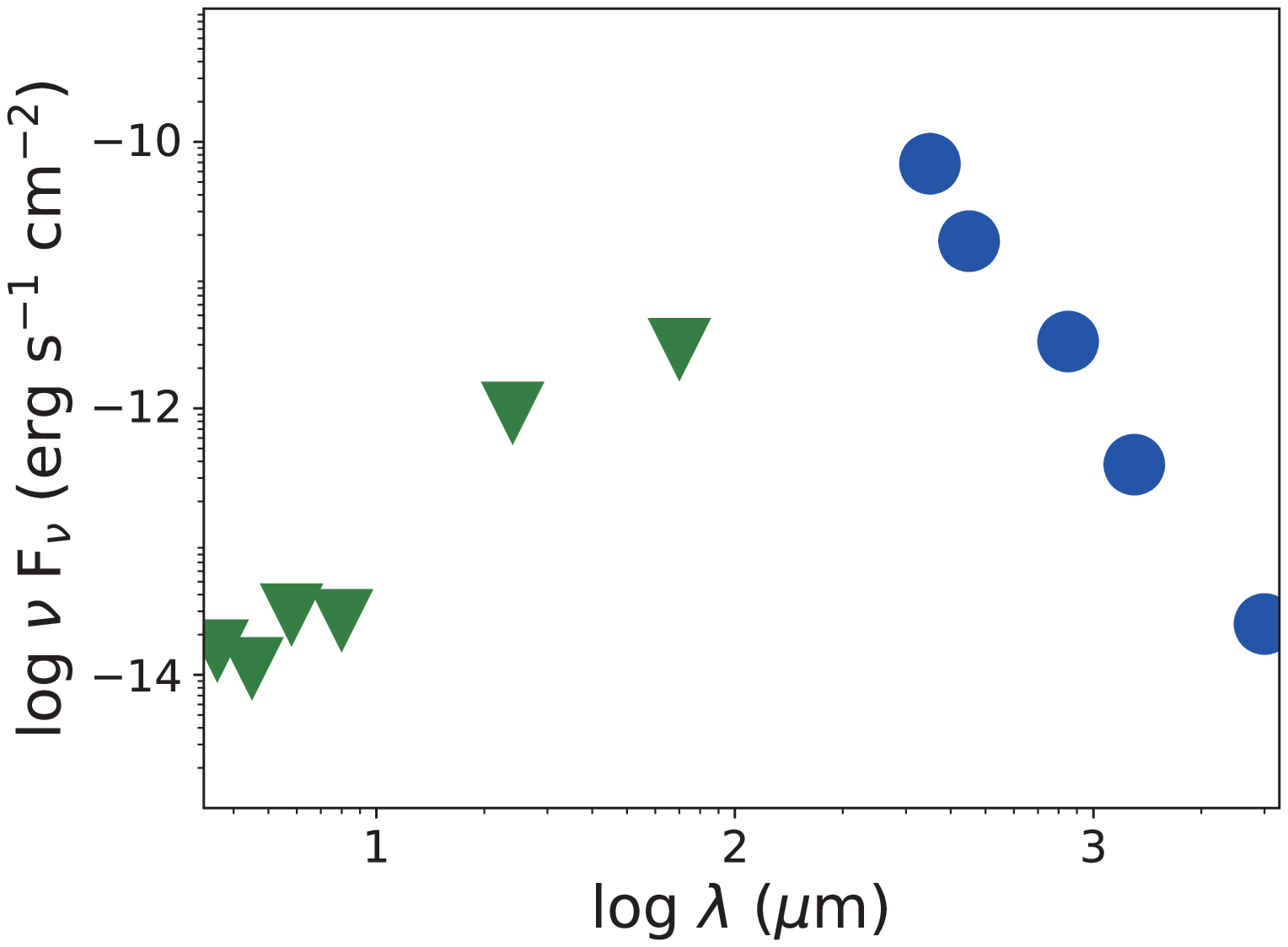}
\caption{SED of SMM11 derived from $Spizter$ IRAC (3.6, 4.5, 5.8, 8.0 $\micron$), MIPS 24 $\micron$, Herschel PACS 70 $\micron$, CSO SHARC-II 350 $\micron$ \citep{su2016}, JCMT SCUBA \citep[450, 850 $\micron$;][]{da1999}, ALMA 1.3 mm (this work), and CARMA 3 mm \citep{leKI2014} data. Blue points denote the measured flux density, where peak intensities in $\JB$ are referred to for SHARC-II and SCUBA wavelengths. Green points denote the $3\sigma$ detection limit for IRAC data, while the upper limits at MIPS and PACS wavelengths are set to be the leaked flux densities from the neighboring protostar (SSTc2d J182959.5$+$011159),
which are higher than 3 times the statistical noise levels.
\label{ch4:fig:sed}}
\end{figure}

To quantify the evolutionary state of SMM11, we measured the flux density at near and mid-infrared wavelengths using $Spitzer$ and $Herschel$. We subtracted average sky levels and then measured the flux densities in apertures twice larger than the FWHMs of the point spread functions (PSF) \citep{an2011}. The bolometric temperature, $T_{\rm bol}$ \citep{my.la1993}, and luminosity $L_{\rm bol}$ are calculated by trapezoidal integration directly from the SED (Figure \ref{ch4:fig:sed}) with other wavelengths in the literature, using $3\sigma$ upper limits for the integration where necessary.
Figure \ref{ch4:fig:11c}c and \ref{ch4:fig:11c}d show that the 24 and 70 $\micron$ emission appear to be contaminated by the Class I source. Nevertheless, the derived bolometric luminosity $L_{\rm bol}\lesssim 0.9\ \Ls$ and particularly bolometric temperature $T_{\rm bol}=$26 K are significantly lower than those of typical protostars \citep[e.g.,][]{kr2012}. The derived bolometric luminosity $0.9\ \Ls$ is an upper limit since its calculation includes the upper limits of flux densities.

\begin{figure}[ht!]
\epsscale{1.1}
\plotone{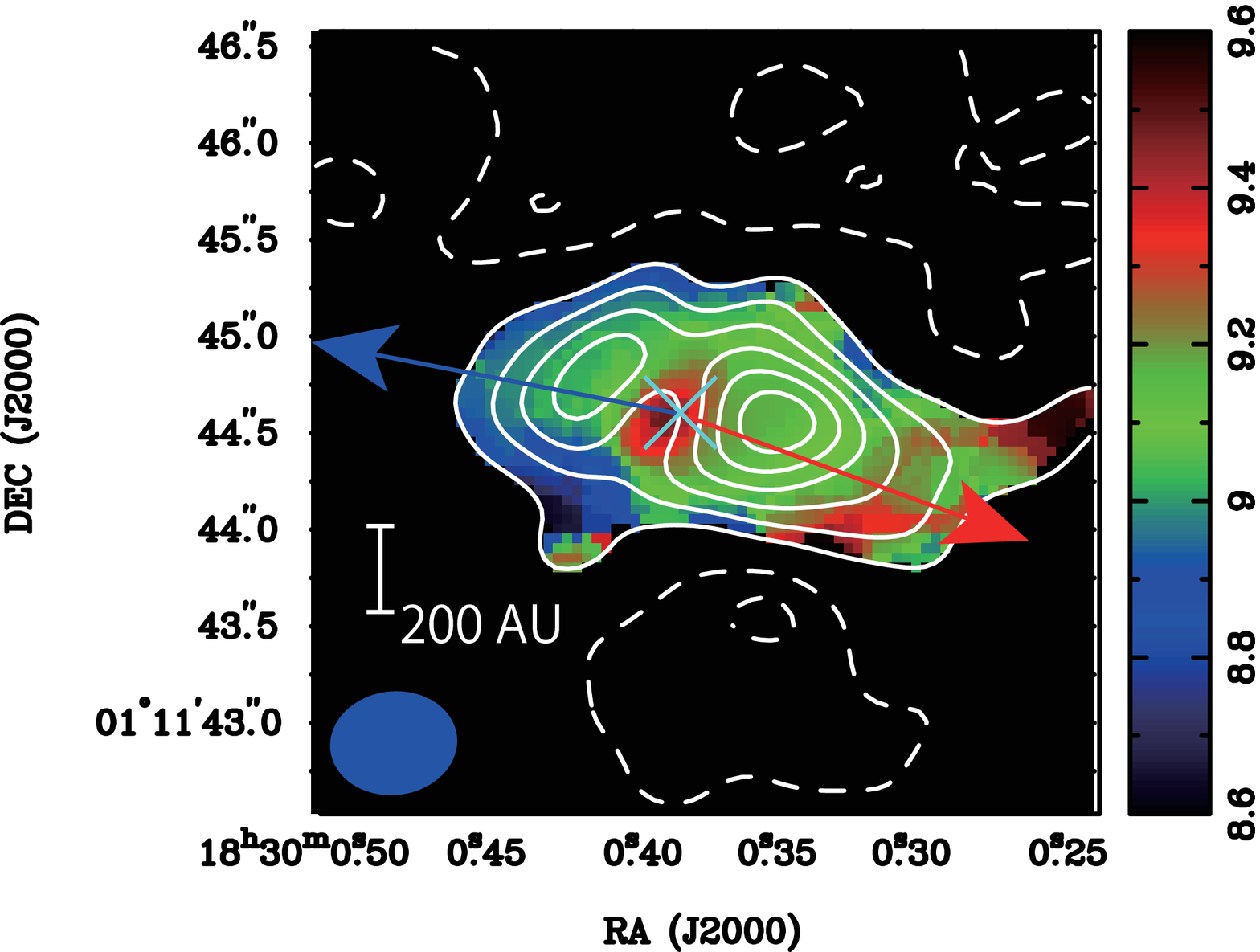}
\caption{Integrated intensity map (white) and mean velocity map (color) of the C$^{18}$O $J=2-1$ emission in SMM11. 
Contour levels are from $3\sigma$ in steps of $3\sigma$, where $1\sigma$ corresponds to $3.5\ \mJB~\kms$. The integrated velocity range of C$^{18}$O emission is from 8.3 to 9.9 $\kms$. X mark shows the 1.3 mm peak position while arrows denote the directions of the $^{12}$CO outflow.
\label{ch4:fig:18mom}}
\end{figure}

Figure \ref{ch4:fig:18mom} shows a map of SMM11 as seen in the C$^{18}$O emission. The emission is elongated along the $^{12}$CO outflow direction, showing a double peak on the eastern and western sides of the continuum peak position. The velocity gradient is overall similar to that of the $^{12}$CO outflow. The systemic velocity of SMM11 and FWHM velocity width of the C$^{18}$O emission are $9.1\ \kms$ and $0.8\ \kms$, respectively, derived from a Gaussian fit to the line profile in a beam area centered on the continuum peak.

\section{DISCUSSION} \label{ch4:sec:dis}
SMM11 is not detected as a point source in 70 $\micron$, 24 $\micron$ (Figure \ref{ch4:fig:11c}), or at shorter wavelengths. Furthermore, its bolometric temperature $T_{\rm bol}=$26 K is in the temperature range theoretically predicted in the FHSC phase \citep{ma1998}. The internal luminosity of SMM11 can also be estimated from its 70 $\mu$m flux density \citep{du2008} to be $L_{\rm int}\lesssim 0.043\ \Ls$, which is also low enough to be consistent with theoretical predictions. 
We now examine whether or not this interpretation is consistent with the other observations.

\subsection{$^{12}$CO outflows} \label{ch4:sec:flow}
The orientation angles of the eastern and western $^{12}$CO lobes were estimated to be P.A.=$79^{\circ}$ and $-110^{\circ}$, respectively, from symmetric axes of the integrated intensity map. Subsequently we fitted the wind-driven-shell model \citep{sh1991,le2000} to the $^{12}$CO integrated intensity map and position-velocity diagrams along the outflow axes as described in Appendix A of \citet{ye2017}. We only used pixels within $10\arcsec$ in radius because the shape of the outflow is more like a bow shock in the outer region and is not well described by the wind-driven-shell model. The fitting implies $c_{0}\sin i=1.3$-$1.9$, $i=77\arcdeg$-$79\arcdeg $, and $v_{0}=3.4$-$4.0\ \kms~{\rm arcsec}^{-1}$ for the eastern $^{12}$CO lobe while $c_{0}\sin i=1.7$-$2.7\ {\rm arcsec}^{-1}$, $i=71\arcdeg$-$87\arcdeg$, and $v_{0}=1.6$-$4.0\ \kms~{\rm arcsec}^{-1}$ for the western $^{12}$CO lobe, where $c_{0}$, $v_{0}$, and $i$ are spatial coefficient ($=z/r^{2}$), velocity coefficient ($v_{r}/r=v_{z}/z$), and inclination angle ($i=0$ means pole-on). The best-fit parabolas are overlaid on Figure \ref{ch4:fig:11c}b, which reproduces the overall shape of the two $^{12}$CO lobes. The inclination angle $i\sim 80\arcdeg$ suggests that the outflow axes lie almost on the plane of the sky. The velocity coefficient $v_{0}\sim 4\ \kms~{\rm arcsec}^{-1}$ suggests a dynamical time, $\sim 600$ yr, and outflow velocity of a few 10 $\kms$ at a few 1000 AU, whereas theoretical simulations in the FHSC phase predict $\sim 5\,\kms$ \citep{ma2008,to2010}.

\subsection{C$^{18}$O abundance} \label{ch4:sec:18}
The C$^{18}$O integrated intensity at the continuum peak position is $30\ \mJB~\kms$ while the continuum peak intensity is $90\ \mJB$. We estimated a fractional abundance of C$^{18}$O relative to H$_{2}$, $X({\rm C}^{18}{\rm O})$, using the continuum peak intensity and a C$^{18}$O Gaussian peak intensity derived from the integrated intensity and the FWHM velocity width $0.8\ \kms$. The dust optical depth is calculated from the continuum peak intensity while that of the dust and C$^{18}$O emitting gas is calculated from the sum of the two peak intensities; the derived values are $\sim$0.1-0.8 when both gas and dust temperatures are 20-50 K with the same opacity and gas/dust ratio as in Section \ref{ch4:sec:results} under the LTE condition.
The derived abundance, $X({\rm C}^{18}{\rm O})$ is $\sim (1.5$-$3.0)\times 10^{-10}$, is more than three orders of magnitude lower than the typical value $5\times 10^{-7}$ in molecular clouds\citep{la1994,wi.ro1994}, suggesting temperatures below CO freeze-out, 20\,K, in the central 100 AU. This is even lower than the overall abundance in Serpens Main \citep{d-c2010} and quantitatively consistent with chemical simulations in the FHSC phase \citep{fu2012,ai2012}. A similar depletion of the carbon-bearing molecule, H$^{13}$CO$^{+}$, is also reported for the FHSC candidates B1-bN and B1-bS by \citet{hu.hi2013}. On the other hand, the elongation and overall velocity gradient of the C$^{18}$O emission suggest molecular desorption on the eastern and western sides which may be due to heating from the associated outflow.

\subsection{Continuum visibility} \label{ch4:sec:vis}
\begin{figure}[ht!]
\epsscale{1.1}
\plotone{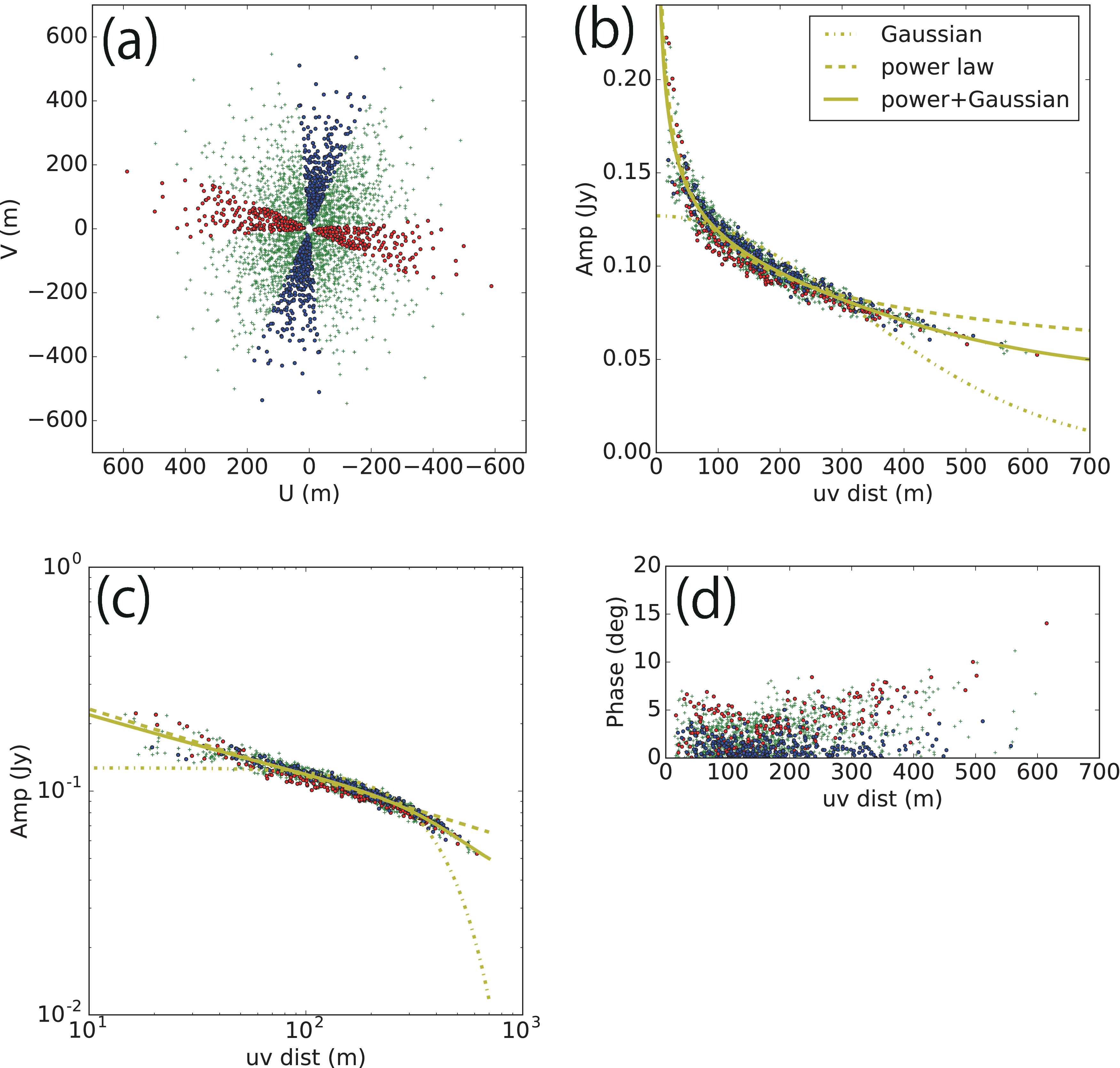}
\caption{Plots of the Continuum visibilities in SMM11. Red and blue circles denote data points at Arctan($U/V)=80^{\circ}\pm 15 ^{\circ}$ and $-10^{\circ}\pm 15^{\circ}$, respectively. (a) Data points on the $uv$-plane. (b) Visibility amplitude as a function of the $uv$-distance. Uncertainty of each amplitude is $\lesssim 1$ mJy. Note that the visibilities also includes contributions from the binary to the northwest of SMM11, 8 mJy. (c) The same plot as (b) but in the log-log plane. (d) Visibility phase as a function of the $uv$-distance. The phase center is set to be the peak position of the continuum image. Dashed-dotted, dashed, and solid curves show the best-fit curves with Gaussian, power-law, and power-law$+$Gaussian functions, respectively.
\label{ch4:fig:uv}}
\end{figure}

Figure \ref{ch4:fig:uv} shows various plots of the continuum visibility data. Each data point corresponds to one baseline and the visibility is averaged over one observational track, $\lesssim 50$ min. Blue and red points denote the visibilities in the major- and minor-axes directions of the continuum image while green points denote those in other directions. In amplitude plots (Figure \ref{ch4:fig:uv}b and c), the blue and red points mostly overlap at $uv$-distance greater than $\sim $40-50 m. This suggests a spherical structure with a radius of $\sim 600$ AU. Most of the phases
are within $\lesssim 5^{\circ}$ from the phase reference center, i.e., the continuum peak position,
(Figure \ref{ch4:fig:uv}d).
To investigate the amplitude distribution in more detail, we fitted it with three different functions: Gaussian, power-law, and a combination of the two. We find that the combination curve $0.031\ {\rm Jy}\ \exp(-\ln 2(\beta/370\ {\rm m})^{2})+0.089\ {\rm Jy}\ (\beta /100\ {\rm m})^{0.33}$ fits the visibility profile better than the other functions, and therefore that the dust structure around SMM11 appears to consist of a compact component with a radius of $\sim 70$ AU and an extended, power-law component. The derived power-law index, $p+q\sim 2.7$, in the outer region ($r\gtrsim 70$ AU), where $p$ and $q$ are the volume density and temperature indices. The Gaussian component has an average H$_2$ number density of $\sim 3\times 10^9\ {\rm cm}^{-3}$ in an inner region ($r\lesssim 70$ AU) if $T_{\rm dust}=30$ K, $\kappa _{850\micron}=0.035\ {\rm cm}^2~{\rm g}^{-1}$, and $\beta=1$.

Similar analyses using continuum visibility are performed for the Class 0/I protostar L1527 IRS by \citet{aso2017}. Its bolometric temperature \citep[44 K;][]{kr2012} suggests that L1527 IRS is more evolved than SMM11. Although L1527 IRS also shows an inclination angle close to edge-on \citep[$\sim 85\arcdeg$;][]{oy2015}, its amplitude distribution shows different profiles along the major- and minor-axes, suggesting a pseudo disk-like envelope, while the envelope of SMM11 is more spherical. These may observationally imply that envelope morphology evolves from spherical shapes into more-disk like shapes.

\subsection{Is SMM11 a FHSC?}
The low $L_{\rm int}$, $T_{\rm bol}$, $X$(C$^{18}$O) values and spherical envelope of SMM11 are consistent with theoretical predictions of the FHSC phase. Although $T_{\rm bol}$ can be weakened by the inclination angle, such an effect is not significant before a disk forms \citep{jo2009}. The bolometric luminosity $\lesssim 0.9\ \Ls$ is much larger than the internal luminosity $0.043\ \Ls$ due to external heating of the envelope around SMM11 and, particularly in this case, contamination from the neighboring protostar. Hence, $L_{\rm int}$ is more directly related with the central heating source. If $L_{\rm int}$ is due to accretion around a protostar, typical parameters indicate a very small protostellar mass, $M_{\rm ps}\lesssim (L_{\rm int}/$0.043 $\Ls)(R/3 \Rs)/(\dot{M}/1\times 10^{-5} \Ms~{\rm yr}^{-1})=4\times 10^{-4} \Ms$ \citep{an2000}. Alternatively, for a typical FHSC radius and mass, the implied mass accretion rate, $\dot{M}\lesssim (L_{\rm int}/$0.043 $\Ls)(R/30 {\rm AU})/(M/0.2 \Ms)=4\times 10^{-5}\ \Ms~{\rm yr}^{-1}$, is consistent with expectations \citep{to2010}.
However, the low luminosity may be due to a low mass accretion rate during episodic accretion and the stellar radius is also uncertain. Moreover, the high velocity of its outflow implies a deeper gravitational potential than predicted for FHSCs. On the other hand, two FHSC candidates reported in previous observations also have high-velocity outflows, such as L1448 IRS2E \citep[$\sim25\ \kms$;][]{ch2010} and B1-bS \citep[$\sim8\ \kms$;][]{hi.li2014}. Furthermore, projection effects and evaluation by characteristic velocities\footnote{Characteristic velocity is defined as the ratio of observed momentum over observed mass, or intensity-weighted mean velocity, and widely used in observational studies about FHSCs \citep[e.g.,][]{du2011} because it is less affected by sensitivity.} could provide lower outflow velocities than the maximum velocity for the other FHSC candidates; in fact, the characteristic outflow velocity of SMM11 is $\sim 5\ \kms$. Hence, FHSC candidates could be divided into two groups, true FHSCs and SMM11-like protostars; the latter suggests criteria with which we can observationally identify an evolutionary phase slightly after the second collapse, where temperature, luminosity, chemistry, and morphology are still similar to those of FHSCs.

Detection limits in various wavelengths are affected by the long distance, 429 pc, of SMM11. For this reason, SMM11 requires additional observations with higher sensitivity and angular resolution than those for other similarly young protostars to better constrain its evolutionary phase and to establish the observational criteria.

\acknowledgments

This paper makes use of the following ALMA data: ADS/JAO.ALMA2015.1.01478.S (P.I. Y. Aso). ALMA is a partnership of ESO (representing its member states), NSF (USA) and NINS (Japan), together with NRC (Canada), NSC and ASIAA (Taiwan), and KASI (Republic of Korea), in cooperation with the Republic of Chile. The Joint ALMA Observatory is operated by ESO, AUI/NRAO and NAOJ.
We thank all the ALMA staff making our observations successful. We also thank the anonymous referee, who gave us invaluable comments to improve the paper.
Data analysis were in part carried out on common use data analysis computer system at the Astronomy Data Center, ADC, of the National Astronomical Observatory of Japan.
Y.A. is supported by the Subaru Telescope Internship Program and acknowledges a grant from the Ministry of Science and Technology (MoST) of Taiwan (MOST 106-2119-M-001-013).
Y.A. acknowledges JSPS KAKENHI Grant Number JP16H00931 in support of this work.
K.S. and M.S. acknowledge JSPS KAKENHI Grant Number JP16K05303 in support of this work
S.T. acknowledges a grant from the Ministry of Science and Technology (MOST) of Taiwan (MOST 102-2119-M-001-012-MY3), and JSPS KAKENHI Grant Number JP16H07086, in support of this work.
K.T. acknowledges JSPS KAKENHI Grant Number JP16H05998 in support of this work.
%

\vspace{5mm}
\facilities{ALMA}


\software{CASA, MIRIAD}

\clearpage
\bibliographystyle{aasjournal}

\begin{thebibliography}{}
\bibitem[Aikawa et al.(2012)]{ai2012} Aikawa, Y., Wakelam, V., Hersant, F., Garrod, R.~T., \& Herbst, E.\ 2012, \apj, 760, 40 
\bibitem[Andre et al.(2000)]{an2000} Andre, P., Ward-Thompson, D., \& Barsony, M.\ 2000, Protostars and Planets IV, 59 
\bibitem[Andrews \& Williams(2005)]{an.wi2005} Andrews, S.~M., \& Williams, J.~P.\ 2005, \apj, 631, 1134 
\bibitem[Aniano et al.(2011)]{an2011} Aniano, G., Draine, B.~T., Gordon, K.~D., \& Sandstrom, K.\ 2011, \pasp, 123, 1218 
\bibitem[Aso et al.(2015)]{aso2015} Aso, Y., Ohashi, N., Saigo, K., et al.\ 2015, \apj, 812, 27 
\bibitem[Aso et al.(2017)]{aso2017} Aso, Y., Ohashi, N., Aikawa, Y., et al.\ 2017, arXiv:1707.08697 
\bibitem[Bate(2011)]{ba2011} Bate, M.~R.\ 2011, \mnras, 417, 2036 
\bibitem[Belloche et al.(2006)]{be2006} Belloche, A., Parise, B., van der Tak, F.~F.~S., et al.\ 2006, \aap, 454, L51 
\bibitem[Boss \& Yorke(1995)]{bo.yo1995} Boss, A.~P., \& Yorke, H.~W.\ 1995, \apjl, 439, L55 
\bibitem[Chen et al.(2012)]{ch2012} Chen, X., Arce, H.~G., Dunham, M.~M., et al.\ 2012, \apj, 751, 89 
\bibitem[Chen et al.(2010)]{ch2010} Chen, X., Arce, H.~G., Zhang, Q., et al.\ 2010, \apj, 715, 1344 
\bibitem[Commer{\c c}on et al.(2012)]{co2012} Commer{\c c}on, B., Levrier, F., Maury, A.~J., Henning, T., \& Launhardt, R.\ 2012, \aap, 548, A39 
\bibitem[Davis et al.(1999)]{da1999} Davis, C.~J., Matthews, H.~E., Ray, T.~P., Dent, W.~R.~F., \& Richer, J.~S.\ 1999, \mnras, 309, 141 
\bibitem[Duarte-Cabral et al.(2010)]{d-c2010} Duarte-Cabral, A., Fuller, G.~A., Peretto, N., et al.\ 2010, \aap, 519, A27 
\bibitem[Dunham et al.(2011)]{du2011} Dunham, M.~M., Chen, X., Arce, H.~G., et al.\ 2011, \apj, 742, 1 
\bibitem[Dunham et al.(2008)]{du2008} Dunham, M.~M., Crapsi, A., Evans, N.~J., II, et al.\ 2008, \apjs, 179, 249-282 
\bibitem[Dunham et al.(2015)]{du2015} Dunham, M.~M., Allen, L.~E., Evans, N.~J., II, et al.\ 2015, \apjs, 220, 11 
\bibitem[Dzib et al.(2011)]{dz2011} Dzib, S., Loinard, L., Mioduszewski, A.~J., et al.\ 2011, Revista Mexicana de Astronomia y Astrofisica Conference Series, 40, 231 
\bibitem[Enoch et al.(2009)]{en2009} Enoch, M.~L., Corder, S., Dunham, M.~M., \& Duch{\^e}ne, G.\ 2009, \apj, 707, 103 
\bibitem[Evans et al.(2009)]{ev2009} Evans, N.~J., II, Dunham, M.~M., J{\o}rgensen, J.~K., et al.\ 2009, \apjs, 181, 321-350 
\bibitem[Furuya et al.(2012)]{fu2012} Furuya, K., Aikawa, Y., Tomida, K., et al.\ 2012, \apj, 758, 86 
\bibitem[Giardino et al.(2007)]{gi2007} Giardino, G., Favata, F., Micela, G., Sciortino, S., \& Winston, E.\ 2007, \aap, 463, 275 
\bibitem[Harsono et al.(2015)]{ha2015} Harsono, D., Bruderer, S., \& van Dishoeck, E.~F.\ 2015, \aap, 582, A41 
\bibitem[Hirano \& Liu(2014)]{hi.li2014} Hirano, N., \& Liu, F.-c.\ 2014, \apj, 789, 50 
\bibitem[Huang \& Hirano(2013)]{hu.hi2013} Huang, Y.-H., \& Hirano, N.\ 2013, \apj, 766, 131 
\bibitem[J{\o}rgensen et al.(2009)]{jo2009} J{\o}rgensen, J.~K., van Dishoeck, E.~F., Visser, R., et al.\ 2009, \aap, 507, 861 
\bibitem[Kristensen et al.(2012)]{kr2012} Kristensen, L.~E., van Dishoeck, E.~F., Bergin, E.~A., et al.\ 2012, \aap, 542, A8 
\bibitem[Lacy et al.(1994)]{la1994} Lacy, J.~H., Knacke, R., Geballe, T.~R., \& Tokunaga, A.~T.\ 1994, \apjl, 428, L69 
\bibitem[Larson(1969)]{la1969} Larson, R.~B.\ 1969, \mnras, 145, 271 
\bibitem[Lee et al.(2000)]{le2000} Lee, C.-F., Mundy, L.~G., Reipurth, B., Ostriker, E.~C., \& Stone, J.~M.\ 2000, \apj, 542, 925 
\bibitem[Lee et al.(2014)]{leKI2014} Lee, K.~I., Fern{\'a}ndez-L{\'o}pez, M., Storm, S., et al.\ 2014, \apj, 797, 76 
\bibitem[Machida et al.(2008)]{ma2008} Machida, M.~N., Inutsuka, S.-i., \& Matsumoto, T.\ 2008, \apj, 676, 1088-1108 
\bibitem[Machida \& Matsumoto(2011)]{ma.ma2011} Machida, M.~N., \& Matsumoto, T.\ 2011, \mnras, 413, 2767 
\bibitem[Masunaga et al.(1998)]{ma1998} Masunaga, H., Miyama, S.~M., \& Inutsuka, S.-i.\ 1998, \apj, 495, 346 
\bibitem[Myers \& Ladd(1993)]{my.la1993} Myers, P.~C., \& Ladd, E.~F.\ 1993, \apjl, 413, L47 
\bibitem[Omukai(2007)]{om2007} Omukai, K.\ 2007, \pasj, 59, 589 
\bibitem[Ortiz-Le{\'o}n et al.(2015)]{or2015} Ortiz-Le{\'o}n, G.~N., Loinard, L., Mioduszewski, A.~J., et al.\ 2015, \apj, 805, 9 
\bibitem[Oya et al.(2015)]{oy2015} Oya, Y., Sakai, N., Lefloch, B., et al.\ 2015, \apj, 812, 59 
\bibitem[Pineda et al.(2011)]{pi2011} Pineda, J.~E., Arce, H.~G., Schnee, S., et al.\ 2011, \apj, 743, 201 
\bibitem[Saigo \& Tomisaka(2006)]{sa.to2006} Saigo, K., \& Tomisaka, K.\ 2006, \apj, 645, 381 
\bibitem[Saigo et al.(2008)]{sa2008} Saigo, K., Tomisaka, K., \& Matsumoto, T.\ 2008, \apj, 674, 997-1014 
\bibitem[Shu et al.(1991)]{sh1991} Shu, F.~H., Ruden, S.~P., Lada, C.~J., \& Lizano, S.\ 1991, \apjl, 370, L31 
\bibitem[Suresh et al.(2016)]{su2016} Suresh, A., Dunham, M.~M., Arce, H.~G., et al.\ 2016, \aj, 152, 36 
\bibitem[Tomida et al.(2010)]{to2010} Tomida, K., Machida, M.~N., Saigo, K., Tomisaka, K., \& Matsumoto, T.\ 2010, \apjl, 725, L239-L244 
\bibitem[Wilner \& Welch(1994)]{wi.we1994} Wilner, D.~J., \& Welch, W.~J.\ 1994, \apj, 427, 898 
\bibitem[Wilson \& Rood(1994)]{wi.ro1994} Wilson, T.~L., \& Rood, R.\ 1994, \araa, 32, 191 
\bibitem[Yen et al.(2017)]{ye2017} Yen, H.-W., Koch, P.~M., Takakuwa, S., et al.\ 2017, \apj, 834, 178 
\end{thebibliography}

\end{document}